# Final Results from and Exploitation Plans for MammoGrid

Chiara DEL FRATE[c], Jose GALVEZ[a], Tamas HAUER[b], David MANSET[d], Richard McCLATCHEY[b], Mohammed ODEH[b], Dmitry ROGULIN[b], Tony SOLOMONIDES[b], Ruth WARREN[e]

[a]CERN, 1211 Geneva 23, Switzerland
[b]CCCS Research Centre, Univ. of West of England, Frenchay, Bristol, BS16 1QY, UK
[c]Istituto di Radiologia, Università degli Studi di Udine, Italy
[d] Maat GKnowledge, Toledo, Spain
[e]Breast Care Unit, Addenbrookes Hospital, Cambridge, UK

**Abstract.** The MammoGrid project has delivered the first deployed instance of a healthgrid for clinical mammography that spans national boundaries. During the last year, the final MammoGrid prototype has undergone a series of rigorous tests undertaken by radiologists in the UK and Italy and this paper draws conclusions from those tests for the benefit of the Healthgrid community. In addition, lessons learned during the lifetime of the project are detailed and recommendations drawn for future health applications using grids. Following the completion of the project, plans have been put in place for the commercialisation of the MammoGrid system and this is also reported in this article. Particular emphasis is placed on the issues surrounding the transition from collaborative research project to a marketable product. This paper concludes by highlighting some of the potential areas of future development and research.

## 1   Introduction

The EU-funded project MammoGrid set out to explore the following conjecture: grid technology and standards have evolved to the point where a prototype federated database of mammograms might be constructed, based on centres in three European countries (UK, Italy and Switzerland). The project was conducted between September 2002 and August 2005 and comprised partners from the universities of Oxford, UWE-Bristol and Cambridge in the UK, CERN in Switzerland and institutes in Pisa, Sassari and Udine in Italy. The project has developed a set of prototypes with a number of regular papers published by MammoGrid partners and the reader is referred to these for the technical aspects of the project in [1], [2], and [3]. In the last year of the project, a final prototype has been constructed and has been deployed to the University hospitals in Cambridge and Udine for proof-of-concept evaluation that would demonstrate the use of a grid-based medical platform in clinical tests.

More specifically, MammoGrid set out to explore the following clinical issues through the delivery of the first grid-based mammography platform:

- *Image standardisation*: The appearance of a mammogram is greatly affected by differences in image acquisition processes (machine type, filter, exposure time etc.). Such differences can impact significantly radiologist judgements (presence of microcalcifications, estimation of proportion of dense breast tissue). Ideally, the given image would be "standardised" by removing such anatomically irrelevant variations prior to adding it to the database. MammoGrid explored the possibility of standardising images using the Standard Mammogram Form$^{TM}$ (SMF) representation developed by Highnam and Brady [4].
- *Breast density as a risk factor*: It has been suggested, primarily by Boyd [5] and others, that the amount/percentage of dense tissue in the breast is a major (perhaps, the major) risk factor for breast cancer (after taking due account of lifetime experiences). The SMF representation provides a number of measures of the amount/proportion of dense tissue, so the MammoGrid clinical partners at Cambridge and Udine hospitals sought to compare measurements of breast density as provided by SMF with the standard methods of visual assessment [6] and the automated 2D interactive computer programme available from Jaffe and Boyd [7].
- *Computer-aided detection of microcalcifications & masses*: Prior to MammoGrid, the project partners from Sassari and Pisa Universities had developed a system named CALMA (Computer Aided Library for Mammography) for the detection of lesions and microcalcifications with reportedly good sensitivity and specificity [8]. The aim was to use the database of mammograms generated during the MammoGrid project at Cambridge and Udine to: (a) re-assess the sensitivity and specificity of CALMA, and (b) to examine whether its performance would be improved on the standardised images generated by SMF.

MammoGrid is one of a number of European Healthgrid projects e.g. [9], [10], and [11]. We note also that the UK "e-Science" programme has funded the e-Diamond project [12], which also aims at developing a federated database of mammograms, but whereas MammoGrid is based on open-source software, eDiamond is based on (IBM) proprietary technology and concentrates on two complementary applications, namely teaching and FindOneLikeIt. Also, in the United States, the National Digital Mammography Archive (NDMA) [13] has adopted a radically different approach, using a large centralised archive.

In this paper, we report on the outcome of the clinical tests in MammoGrid and identify how far the project progressed towards real clinical use of a healthgrid. The paper identifies both achievements and obstacles in the use of a deployed grid-based healthcare application and it also highlights the lessons learned from MammoGrid. In addition, the exploitation plans for MammoGrid are outlined after the post-project era has begun for commercialisation of the project software.

## 2   The MammoGrid Clinical Evaluation

The MammoGrid project has several technical aspects, which are briefly mentioned in this section in order to convey a flavour of the scope and complexity of the challenges that were met. These aspects include: image standardisation using SMF; the development of a workstation on which images can be acquired, annotated, and uploaded to the grid; and the distribution of data, images and clinician queries across grid-based databases, while respecting ethical, legal, confidentiality and security constraints applicable differently in the partner countries of origin. It was not the intention of this project to produce new grid software. Rather, the aim has been to use, wherever possible, open source software to provide middleware services to enable radiologists to query patient records across a widely distributed "federated" database of mammographic images and to perform epidemiological and computer-aided detection on the sets of returned images. For example, in the MammoGrid project, radiologists may annotate (i.e. mark out) different regions of a mammogram, which are then subjected to different computer-aided detection algorithms (including CALMA) and compared with stored mammograms in the database. Since any one of these stages may be executed independently or take some time to be completed, the process must be controlled in a way that recognises the current state of the computation and ensures that results are meaningfully assembled from the various partial outcomes. To provide for these possibilities, MammoGrid adopted AliEn (Alice Environment [15]) a light-weight grid middleware developed to satisfy the needs of the ALICE experiment at CERN for large scale distributed computing. The details of AliEn and how it was used in the design of MammoGrid are beyond the scope of this paper (c.f. [2]).

The MammoGrid project has delivered its final proof-of-concept prototype enabling clinicians to store digitised mammograms along with appropriately anonymized patient metadata; the prototype provides controlled access to mammograms both locally and remotely stored. A typical database comprising several thousand mammograms has been created for user tests of clinicians' queries. The prototype comprises

- a high-quality clinician visualisation workstation (used for data acquisition and inspection);
- an imaging standard-compliant interface to a set of medical services (annotation, security, image analysis, data storage and querying services) residing on a so-called 'Grid-box'; and
- secure access to a network of other Grid-boxes connected through grid middleware.

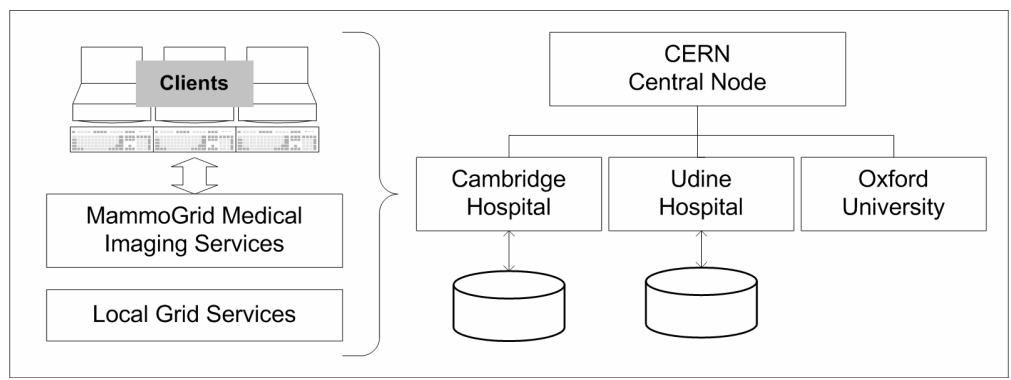

**Figure 1:** The MammoGrid Virtual Organisation

To facilitate evaluation of the final prototype at the clinical sites, a MammoGrid Virtual Organisation (MGVO) was established and deployed (as shown in figure 1). The MGVO is composed of three mammography centres – Addenbrookes Hospital, Udine Hospital, and Oxford University. These centres are autonomous and independent of each other with respect to their local data management and ownership. The Addenbrookes and Udine hospitals have locally managed databases of mammograms, with several thousand cases between them. As part of the MGVO, registered clinicians have access to (suitably anonymized) mammograms, results, diagnosis and imaging software from other centres. Access is coordinated by the MGVO central node at CERN.

The service-oriented architecture approach (SOA) adopted in MammoGrid permits the interconnection of communicating entities, called services, which provide functionality through exchange of messages. The services are 'orchestrated' in terms of service interactions: how services are discovered, how they are invoked, what can be invoked, the sequence of service invocations, and who can execute them.

The MammoGrid Services (MGs) are a set of services for managing mammographic images and associated patient data on the grid. Figure 2 illustrates the services that make up the MGVO, (For simplicity, the node at Oxford University has not been included). The MGs are: (a) Add for uploading files (DICOM [16] images and structured reports) to the MGVO; (b) Retrieve for downloading files from the grid system; (c) Query for querying the federated database of mammograms; (d) AddAlgorithm for uploading executable code to the grid system; (e) ExecuteAlgorithm for executing grid-resident executable code on grid-resident files on the grid system; and (f) Authenticate for logging into the MGVO. For further details consult [3].

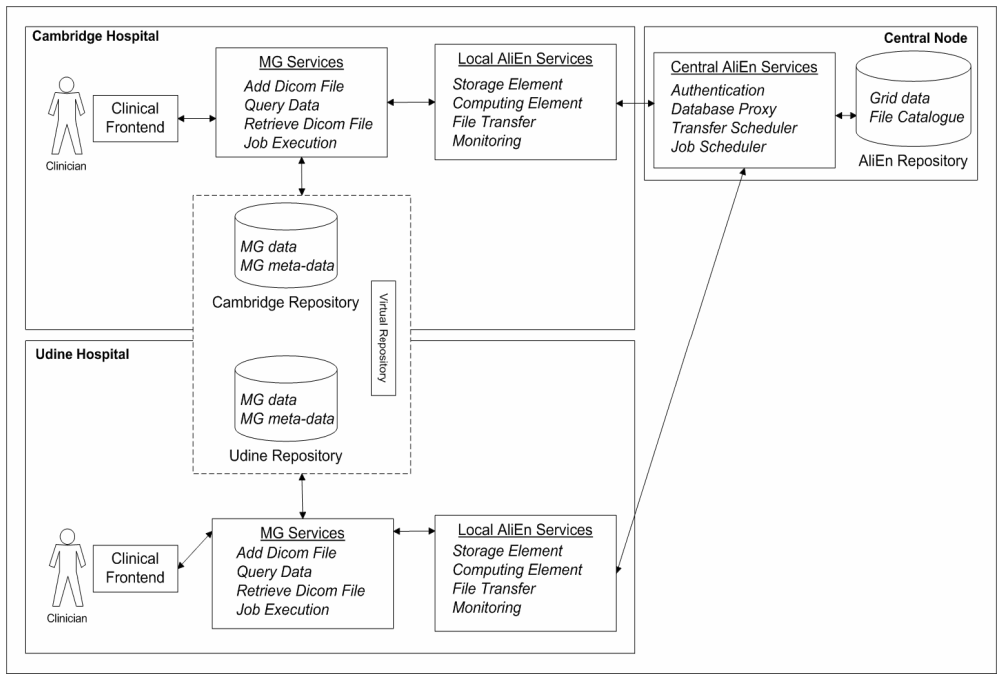

**Figure 2:** The MammoGrid Services in the MGVO.

Currently the MGVO encompasses data accessible to senior radiologists at the Addenbrookes and Udine hospitals, as well as researchers at Oxford University. The radiologists have been able to view raw image data from each other's hospitals and have been able to second-read grid-resident mammograms and each to annotate separately the images for combined diagnosis. This has demonstrated the viability of distributed image analysis using the grid and shown considerable promise for future grid-based health applications. Despite the anticipated performance limitations that existing grid software and networks impose on system usage, the clinicians have been able to discover new ways to collaborate using the virtual organization. These include the ability to perform queries over a virtual repository spanning data held in Addenbrookes and Udine.

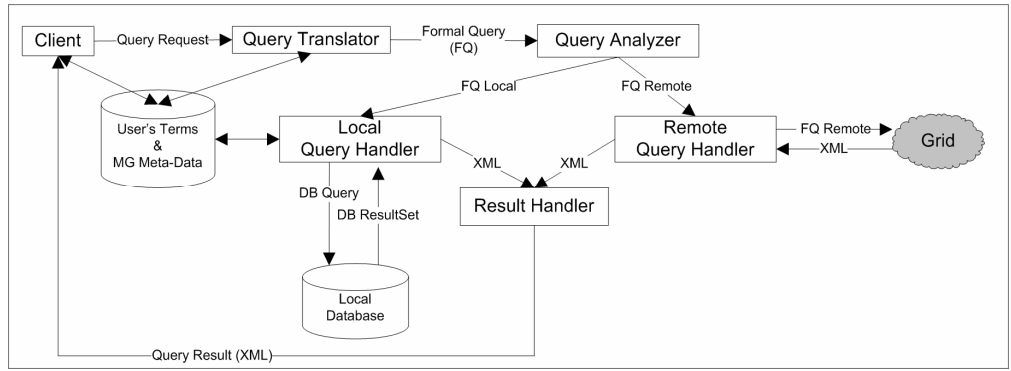

**Figure 3:** Clinical Query Handing in MammoGrid.

Clinicians define their mammogram analysis in terms of queries they wish to be resolved across the collection of data repositories. Queries can be categorised into simple queries (mainly against associated data stored in the database as simple attributes) and complex queries which require derived data to be interrogated or an algorithm to be executed on a (sub-)set of distributed images. The important aspect is that image and data distribution are transparent for radiologists so that queries can be formulated and executed as if these records were locally resident. Queries are executed at the location where the relevant data resides, i.e. sub-queries are moved to the data, rather than large quantities of data being moved to the clinician, which can be prohibitively expensive given the volume especially of image data. Figure 3 illustrates how queries are handled in MammoGrid.

The Query Analyzer takes a formal query representation and decomposes it into (a) a formal query for local processing, and (b) a formal query for remote processing. It then forwards these decomposed queries to the Local Query Handler and the appropriate Remote Query Handler for the resolution of the request. The Local Query Handler generates query language statements (e.g. SQL) in the language of the associated Local DBMS (e.g. MySQL). The result set is converted to XML and routed to the Result Handler. The Remote Query Handler is a portal for propagating queries and results between sites. This handler forwards the formal query for remote processing to the Query Analyzer of the remote site. Finally the remote query result set is converted to XML and routed to the Result Handler. As of writing the MGVO holds:

| Site | Number of Patients | Total Number of Image Files | Number of SMF Files | Associated Image Data Size | File Storage Size |
|---|---|---|---|---|---|
| Cambridge | 1423 | 9716 | 4815 | 14 Mb | 260 Gb |
| Udine | 1479 | 17285 | 8634 | 23.5 Mb | 220 Gb |

**Table 1:** Virtual Repository size of the MammoGrid prototype.

The average processing time for the core services are: (1) Add 8Mb DICOM file takes approximately 7 seconds, (2) Retrieve 8Mb DICOM file from a remote site takes approximately 14 seconds, and (3) the SMF workflow of ExecuteAlgorithm takes around 200 seconds. Table 2 presents examples of queries and their execution results.

| Query | Cambridge | Udine | Num images | Num patients |
|---|---|---|---|---|
| By Id: Cambridge patient | 2.654 sec | 2.563 sec | 8 | 1 |
| By Id: Udine patient | 2.844 | 3.225 | 16 | 1 |
| All female | 103 | 91 | 12571 | 1510 |
| Age [50,60] and ImageLaterality=L | 19.489 | 22.673 | 1764 | 357 |

**Table 2**: Data query performance of the MammoGrid prototype.

In the final months of the project, clinicians have been testing the MammoGrid prototype functionality across two clinical studies. First, the Standard Mammogram Format (SMF) [4] software has been used to measure breast density. This clinical phase of the project, designed jointly by Cambridge and Udine explored the relationship between mammographic density, age, breast size, and radiation dose. In this phase, breast density has been measured by SMF and compared with standard methods of visual assessment. Heights, weights, and mass indicators were used in an international comparison, and the output demonstrated a richer dataset would be needed to study effects of lifestyle factors such as diet or HRT use between the two national populations.

Second, the University of Udine led a project to validate the use of SMF in association with Computer Aided Detection (CADe) from the CALMA project [17] & [18]. Cancers and benign lesions were supplied from the clinical services of Udine and Cambridge to provide the benchmarking and the set of test cases. Cancer cases included women whose unaffected breast served the density study to provide cases for the CADe analysis from the affected side mammogram. Beyond the relevant clinical results obtained from these studies, the MammoGrid project has shown that these new forms of clinical collaboration can be supported using the grid (see [19] and [20]).

## 3     Lessons Learned

The nature of the project and its particular constraints of multi-disciplinarity, dispersed geographical development, large discrepancies in participants' domain knowledge whether of software engineering techniques or of breast cancer screening practice, as well as the novelty of the grid environment, provide experiences from which other grid-based medical informatics projects can benefit. We summarize below some of the main lessons that can be learned in this context.

First, the project was particularly fortunate in its medical partners. In general, the medical environment is highly risk-averse, conservative in nature and reluctant to adopt new technologies without significant evidence of tangible benefit. It is therefore important to identify a suitable user community in which new technologies (such as grid) can be evaluated. In the case of MammoGrid we have had real commitment from the radiology community in the project's requirements definition, analysis, implementation, and evaluation and this was crucial to the success of the project. The data samples used were of a sensitive nature and required both ethical clearance from participating institutions and anonymization of the data and even then only for strictly research use in the project. Many ethical/legal obstacles remain to be tackled before clinicians can share sensitive patient data between institutes, never mind across national boundaries.

Second, it has become clear from our experiences that grid middleware technology itself is still evolving, and this suggests that there is a clear need for standardization to enable production-quality systems to be developed. Despite the availability of toolsets such as the Globus 4.0 [21], the development of applications that harness the power of the grid at present requires specialist skills and is thus still costly in terms of manpower. Only with the arrival of stable middleware and packaged grid services will the development of medical applications become viable.

Third, the performance of existing middleware is also somewhat suspect; the MammoGrid project had to circumvent some of the delivered grid services to ensure adequate system performance for its prototype evaluation. For example, the database of medical images was completely decoupled from the grid software to provide adequate response for MammoGrid query handling. The EGEE [14] project is addressing these technological deficiencies and improved performance of the middleware should consequently be delivered in the coming years.

Fourth, grid technology for medical informatics is still in its infancy and needs some proven examples of its applicability; MammoGrid is the first such exemplar in practice. Equally, awareness of grid technology and its potential (and current limitations) must still be raised in the target user communities such as Health, Biomedicine, and more generally life sciences.

Fifth, the project has indicated that it is possible to use modelling techniques (such as use-cases from UML) in a widely distributed, multi-disciplinary software engineering problem domain, provided a very pragmatic approach is used, where adopting a certain modelling technique is, to some extent, independent from the software development life cycle model being applied. The MammoGrid project has benefited significantly in its coordination, communication and commitment by utilizing the use-case model as the lingua franca during user requirement analysis and system design rather than following the disciplines of the Rational Unified Process (RUP) to the letter; see [22] for a detailed account of this approach.. Furthermore, this has also demonstrated the possibility of bridging the gap between use-case models and grid-based service oriented architectures as demonstrated by the transition from the MammoGrid's use-case model to its respective SOA architectural model.

Sixth, the evolutionary approach to system development work packages has mitigated the effects of the project constraints of a highly dynamic research-oriented environment in which novices and specialists in software engineering have worked together even though they may have been geographically separated.

Further areas that might promote the use of rigorous software engineering disciplines in the design of grid-based software services are that of model-driven engineering [23] and the use of architecture descriptions ([24] and [25]) as the basis for the generation of grid-wide services. These aspects are, however, outside the scope of the current project.

## 4    Future Exploitation Plans

By using grid computing, the MammoGrid system allows hospitals, healthcare workers and researchers to share data and resources, while benefiting from an augmented overall infrastructure. It supports effective co-working, such as obtaining second opinion, provides the means for powerful comparative analysis of mammograms and opens the door to novel, broad-based statistical analysis of the incidence and forms of breast cancer. Through the MammoGrid project, partners have developed a strong collaboration between radiologists active in breast cancer research and academic computer scientists with expertise in the applications of grid computing. The success of the project has led to interest from outside companies and hospitals, with one Spanish company, Maat GKnowledge, looking to deploy a commercial variant of the system in three hospitals of the Extremadura region in Spain. Maat GKnowledge aims to provide doctors with the ability to verify test results, to obtain a second opinion and to make use of the clinical experience acquired by the hospitals involved in the project. They then aim to scale the system up and to expand it to other areas of Spain and then Europe. With the inclusion of new hospitals, the database will increase in coverage and the knowledge will increase in relevance and accuracy, enabling larger and more refined epidemiological studies. Therefore, clinicians will be provided with a significant data set to serve better their investigations in the domain of cancer prevention, prediction and diagnoses. This will result in improved research quality as well as improved citizen access to the latest healthcare technologies.

The MammoGrid system prototype has been at the leading edge of the healthgrid revolution and implements for the first time such a solution for mammogram acquisition and manipulation. The resulting application has reached a high level of complexity which now requires continued partnership between academics, clinicians and industry to provide the necessary technology transfer and to enable real commercialisation. In this context, the MammoGrid Technology Transfer and Innovation eXchange (MaTTrIX) project has been proposed as an ideal means to transfer the project knowledge and expertise from the research to the commercial domain, to make its innovation available to the company Maat GKnowledge in Spain, to carry forward research findings to radiological practice and to reinforce existing partnerships between networks of clinicians and technologists. To achieve these objectives, the intention is to introduce a service which exploits the findings of the MammoGrid project in practice, while the researchers also learn from the application of their ideas in a real environment. This will create a two-way innovation flow between the academic and commercial worlds, building on existing synergies and collaborations, and improving the overall viability and commercialisation of the MammoGrid system software.

To this end, the host organisation Maat GKnowledge, the University of the West of England, Bristol, CIEMAT (Centro de Investigaciones Energéticas, Medioambientales y Tecnológicas), Spain, and CERN together with clinical partners at the University hospitals in Cambridge and Udine and at the Hospitals Infanta Cristina, Merida and San Benito in Extremadura are initiating a transfer of knowledge, research competences and technologies to enable the company to take over future development of the MammoGrid system. This will enable the training of existing Maat GKnowledge staff in MammoGrid technologies and the acquisition of the technology and know-how for commercialisation. This transfer would constitute the mechanism for the creation and development of a durable technology and knowledge transfer partnership.

The MaTTrIX technology transfer approach relies on two main cornerstones:

- to promote the mobility of key experienced researchers, to absorb, expand and disseminate the knowledge for the MammoGrid system to evolve gradually into a commercial offering, making it available for healthcare across the European Research Area (ERA) as a viable and clinically assessed solution; and

- to create and to develop a strategic and durable partnership between new hospitals and partners of the MammoGrid project, providing a sound foundation for a network of excellence in European research.

Having paved the way for potential knowledge discovery in the understanding of breast cancer, the MammoGrid project has also determined new important research paths for better cancer prediction and diagnosis. The use of a standard format for mammogram images (SMF) and its outcome in the epidemiological investigations, has demonstrated the relevance of new grid-based clinical studies, which may lead to major advances in cancer prediction. In addition, while most computer-aided detection (CADe) systems process raw and noisy data at the price of accuracy and quality, the implemented solutions in MammoGrid have indicated the valuable joint use of SMF and CADe tools to improve automated cancer diagnosis. Considering the IT contribution in MammoGrid, the distributed technologies used in the project combined with the recent clinical feedback, obtained from the assessment of the proof-of-concept prototype, have highlighted the importance of offering a collaborative platform for healthcare. Not only would such a solution demonstrate the benefits of clinical second opinion, but it would also help in reducing the information infrastructure costs by enabling heterogeneous and scalable resource sharing, resulting in an improved system access even to less favoured regions and countries in Europe.

Emphasising this last point the Hospitals of Infanta Cristina, Merida and San Benito in Extremadura in Spain, will obtain access to the MammoGrid system and expertise at a reduced cost, the infrastructure being already in place. While accessing the latest technologies and software, these hospitals will share and enrich their clinical experience by interacting with other trained clinicians from the university hospitals in Cambridge and Udine to obtain expertise in the use of SMF and CADe. Academic computer scientists will continue to analyse the ways in which new functionality is exploited by radiologists in their protocols and workflows, maintaining the design of the system under review. This new partnership will result in improved processes locally and also in refined clinical knowledge being made available in a Europe-wide production reference database, enabling new clinical studies in the spirit of improving cancer prediction and detection. It is hoped that this collaboration will provide a significant exemplar for the European Research Area.

## 5    Conclusions

The MammoGrid project has deployed its first prototype and has performed the first phase of in-house tests, in which a representative set of mammograms have been tested between sites in the UK, Switzerland and Italy. In the next phase of testing, clinicians will be closely involved in performing tests and their feedback will be reflectively utilised in improving the applicability and performance of the system. In its first two years, the MammoGrid project has faced interesting challenges originating from the interplay between medical and computer sciences, and has witnessed the excitement of the user community whose expectations from a new paradigm are understandably high. As the MammoGrid project moves into its final implementation and testing phase, further challenges are anticipated. In conclusion, this paper has outlined the MammoGrid application's deployment strategy and experiences.  Also, it outlines the strategy being adopted for migration to the new lightweight middleware called gLite [26].

MammoGrid is one of several current projects that aim to harness recent technological advances to achieve the goal of complex data storage in support of medical applications. The approaches vary widely, but at least two projects, e-Diamond in the UK and MammoGrid in Europe, have adopted the grid as their platform of choice for the delivery of improved quality control, more accurate diagnosis and statistically significant epidemiology. Since breast screening in the UK and in Italy has been based on film, mammograms have had to be digitised for use in both e-Diamond and MammoGrid.  By contrast, the NDMA project in the United States has opted for centralised storage of direct digital mammograms. The next step for MammoGrid, its application in the Spanish region of Extremadura, will be based on both film and direct digital mammography.

The central feature of the MammoGrid project is a geographically distributed, grid-based database of standardised images and associated patient data. The novelty of the MammoGrid approach lies in the application of grid technology and in the provision of data and tools, which enable radiologists to compare new mammograms with existing ones on the grid database, allowing them to make comparative diagnoses as well as judgements about quality. In the longer term, as the potential of the database is to be populated with provenance-controlled, reliable data from across Europe, with the prospect of statistically robust epidemiology that allows analysis of 'lifestyle' factors, including, e.g., diet, exercise and exogenous hormone use. And, hence the grid would also be suitable for storing genetic or pathological image information.

The project has attracted attention as a paradigm for grid-based radiology and imaging applications. While it has not yet solved all problems, the project has established an approach and a prototype platform sharing medical data, especially images, across a grid. In loose collaboration with a number of other European medical grid projects, it is addressing the issues of informed consent and ethical approval, data protection, compliance with institutional, national and European regulations, and security [27], [28]. In conclusion, the MammoGrid project may be considered as a major advance in bridging the gap between the grid as an advanced distributing computing infrastructure and the medical domain and therefore should enable further grid-based projects to benefit from both its main lessons and its results.

**Acknowledgements**

The authors thank the European Commission and their institutes for support and acknowledge the contribution of the following MammoGrid collaboration members: Mike Brady and Chris Tromans (University of Oxford), Predrag Buncic, Pablo Saiz (CERN/AliEn), Martin Cordell, Tom Reading and Ralph Highnam (Mirada) Piernicola Oliva (University of Sassari) and Evelina Fantacchi and Alessandra Rettico (University of Pisa). The assistance from the MammoGrid clinical community is warmly acknowledged especially that from Iqbal Warsi and Jane Ding of Addenbrookes Hospital, Cambridge, UK and Dr Massimo Bazzocchi of the Istituto di Radiologia at the Università degli Studi di Udine, Italy. Last, but by no means least, the authors are indebted to the former MammoGrid Project Coordinator, Roberto Amendolia, both for his original innovative contribution and his robust support.